\documentclass[epjCONF]{svjour}
\usepackage{graphics}
\usepackage[varg]{txfonts} 
\usepackage[latin1]{inputenc}
\session-title{FUSION11}

\newcommand{\al}{\alpha}

\newcommand{\az}{\varphi}
\newcommand{\ro}{\rho}

\newcommand{\be}{\beta}
\newcommand{\ga}{\gamma}

\newcommand{\oeq}{\begin{equation}}
\newcommand{\ceq}{\end{equation}}
\newcommand{\oeqn}{\begin{eqnarray}}
\newcommand{\ceqn}{\end{eqnarray}}

\renewcommand{\>}{\rangle}
\newcommand{\<}{\langle}
\renewcommand{\(}{\left(}
\renewcommand{\)}{\right)}
\renewcommand{\[}{\left[}
\renewcommand{\]}{\right]}

\newcommand{\stf}{\,\,\,}

\newcommand{\stb}{\!\!\!}

\newcommand{\kfi}{|\phi \>}

\newcommand{\kvac}{|-\>}

\newcommand{\bfi}{\<\phi |}

\newcommand{\oX}{\hat{X}}
\newcommand{\oY}{\hat{Y}}

\newcommand{\oH}{\hat{H}}

\newcommand{\oD}{\hat{D}}

\newcommand{\oro}{\hat{\rho}}

\newcommand{\oad}{\hat{a}^\dagger}
\newcommand{\oa}{\hat{a}}

\newcommand{\oB}{\hat{B}}

\newcommand{\oq}{\hat{q}}

\renewcommand{\d}{{\mbox d}}

\newcommand{\Tr}{\mbox{Tr}}
\newcommand{\tr}{\mbox{tr}}

\begin{document}
\title{Actinide collisions for QED and superheavy elements with the time-dependent Hartree-Fock theory and the Balian-V\'en\'eroni variational principle}
\author{C\'edric Simenel\inst{1,2}\fnmsep\thanks{\email{cedric.simenel@cea.fr}} \and C\'edric Golabek\inst{3} \and David J. Kedziora\inst{2} }
\institute{CEA, Centre de Saclay, IRFU/Service de Physique Nucl\'eaire, F-91191 Gif-sur-Yvette, France \and Department of Nuclear Physics, Research School of Physics and Engineering, Australian National University, Canberra, Australian Capital Territory 0200, Australia \and GANIL (IN2P3/CNRS - DSM/CEA), BP 55027, F-14076 Caen Cedex 5, France.}
\abstract{
Collisions of actinide nuclei form, during very short times of few zs ($10^{-21}$~s),
the heaviest ensembles of interacting nucleons available on Earth.
Such collisions are used to produce super-strong electric fields by the huge number of interacting protons 
to test spontaneous positron-electron pair emission (vacuum decay) predicted by the quantum electrodynamics (QED) theory. 
Multi-nucleon transfer in actinide collisions could also be used as an alternative way to fusion 
in order to produce neutron-rich heavy and superheavy elements thanks to inverse quasifission mechanisms.
Actinide collisions are studied in a dynamical quantum microscopic approach.
The three-dimensional time-dependent Hartree-Fock (TDHF) code {\textsc{tdhf3d}} is used 
with a full Skyrme energy density functional to investigate the time evolution 
of expectation values of one-body operators, such as fragment position and particle number.
This code is also used to compute the dispersion of the particle numbers 
(e.g., widths of fragment mass and charge distributions) from TDHF transfer probabilities,
on the one hand, and using the Balian-Veneroni variational principle, on the other hand. 
A first application to test QED is discussed.
Collision times in $^{238}$U+$^{238}$U are computed to determine the optimum energy for the observation of the vacuum decay. 
It is shown that the initial orientation strongly affects the collision times and reaction mechanism. 
The highest collision times predicted by TDHF in this reaction are of the order of 
$\sim4$~zs at a center of mass energy of 1200~MeV.
According to modern calculations based on the Dirac equation,
the collision times at $E_{cm}>1$~GeV are sufficient to allow spontaneous electron-positron pair emission from QED vacuum decay, 
in case of bare uranium ion collision. 
A second application of actinide collisions to produce neutron-rich transfermiums is discussed. 
A new inverse quasifission mechanism associated to a specific orientation of the nuclei is proposed 
to produce transfermium nuclei ($Z>100$) in the collision of prolate deformed actinides 
such as $^{232}$Th+$^{250}$Cf. 
The collision of the tip of one nucleus with the side of the other results in a nucleon flux toward the latter.
The probability distributions for transfermium production in such a collision are computed.
The produced nuclei are more neutron-rich than those formed in fusion reactions, 
thus, leading to more stable isotopes closer to the predicted superheavy island of stability.
In addition to mass and charge dispersion, the Balian-Veneroni variational principle
is used to compute correlations between $Z$ and $N$ distributions, which are zero in standard TDHF calculations.} 
\maketitle

\section{Introduction}
\label{intro}

Actinide collisions are  important tools to test our understanding of the nuclear many-body problem. 
They form nuclear systems in extreme conditions of mass and isospins. 
The prediction of the outcome of such collisions is a great challenge for nuclear theorists. 

In particular, the question of ''How long can two actinides stick together'' is of wide interests. 
The quantum-electrodynamic (QED) theory predicts that spontaneous pairs of $e^++e^-$ may be emitted due to the strong electric fields produced by the protons~\cite{rei81,gre83,ack08}. This process is also known as ''QED vacuum decay''. It occurs when an empty electron state dives into the Dirac sea. 
QED predicts that such a hole state is unstable and decays by  $e^++e^-$ pair production. The life-time for such process is, however, longer than the collision-time between actinides. Then, the latter has to be optimized to allow for an experimental observation of vacuum decay. 
Recent 
calculations based on the time-dependent Dirac equation~\cite{ack08} show that two bare $^{238}$U need to stick together during at least $2$~zs to allow for an observation of spontaneous positron emission. 
Although no pocket exists in the nucleus-nucleus potential of this system~\cite{kat89,ber90,tia08,sim09}, nuclear attraction reduces Coulomb repulsion and dissipation mechanisms such as evolution of nuclear shapes may delay the separation of the system~\cite{zag06}. 
In a recent experiment, delay times in this reaction was searched analyzing kinetic energy loss and mass transfer~\cite{gol10}. 

Another application of actinide collisions is to form neutron-rich heavy and superheavy nuclei by multi-nucleon transfer~\cite{vol78,fre79,zag06}. 
Such reactions could be used to explore the ''blank spot'' between decay chains of nuclei formed by ''hot'' and ''cold fusion'' around $Z=105$ and $N=160$. 

Theoretically, the complexity of reaction mechanisms and the high number of degrees of freedom to be included motivate the use of microscopic approaches.  
Early dynamical microscopic calculations of $^{238}$U+$^{238}$U were performed with spatial symmetries and simplified effective interactions~\cite{cus80,str83}.
Recently, this system has been studied within the Quantum Molecular Dynamics (QMD) model~\cite{tia08,zha09} in which nucleon wave functions are constrained to be Gaussian wave packets and with the time-dependent Hartree-Fock approach which overcomes this limitation~\cite{gol09}. 

\begin{figure*}
\resizebox{2.0\columnwidth}{!}{
\includegraphics{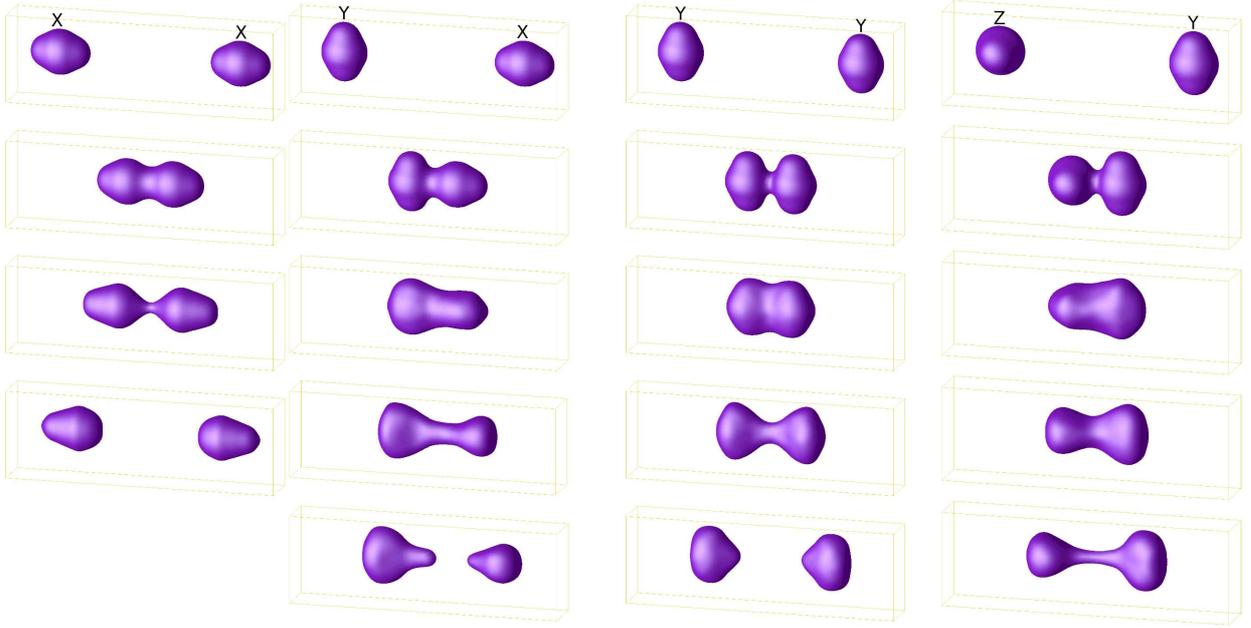} }
\caption{Isodensities at half the saturation density, i.e., $\rho_0/2=0.08$~fm$^{-3}$, in $^{238}$U+$^{238}$U central collisions at a center of mass energy $E_{c.m.}=1200$~MeV. Evolutions associated to the four initial configurations $xx$, $yx$, $yy$, and $zy$ are plotted in columns (time runs from top to bottom). Consecutive snapshots are separated by 1.125~zs. }
\label{fig:densities}
\end{figure*}

\section{Formalism}

In general, the full quantum many-body problem cannot be solved exactly and,
in most realistic cases, approximations have to be made. 
In general, variational principles are useful to build approximation schemes by reducing the variational space. 

\subsection{The Balian-V\'en\'eroni variational principle}

The Balian and V\'en\'eroni (BV) variational principle is
based on the action~\cite{bal81}
\oeq
S_{BV}=\Tr \[\oD(t_1)\oB(t_1)\]-\int_{t_0}^{t_1} \stb \d t\stf \Tr\(\oB\frac{\partial\oD}{\partial t}-i\oD[\oH,\oB]\),
\ceq
where $\oB$ and $\oD$ are the time-dependent trial observable and density matrix of the trial state, respectively, 
and $\Tr$ denotes a trace in the Fock space. Both the state and the observable are allowed to vary between $t_0$ and $t_1$,
corresponding to a mixture of the Schr\"odinger and Heisenberg pictures. 
They are constrained to obey the mixed boundary conditions $\oD(t_0)=\oD_0$ and $\oB(t_1)=\oX$, 
where $\oD_0$ is the density matrix of the initial state of the system and $\oX$ is the operator  
we want to evaluate at time $t_1$. 
Without restriction on the variational spaces, the variational principle $\delta{S}_{BV}=0$, 
with the above conditions, is fully equivalent to the Schr\"odinger equation if the initial state is pure ($\oD_0=|\Psi_0\>\<\Psi_0|$).

\subsection{Mean-field approximation}

In most practical applications, mean-field models are considered in a first approximation. 
In mean-field theories, the interaction between the particles is replaced 
by a one-body mean-field potential generated by all the particles. 
It is then assumed that each particle evolves independently in this potential. 

For instance, $N$ independent fermions may be described by a 
Slater determinant~$\kfi=\prod_{i=1}^N\oad_i\kvac$, where $\oad_i$ creates a particle in the state $|\az_i\>$ 
when it is applied to the particle vacuum~$\kvac$.
In such a Slater determinant, all the information is contained in the one-body density-matrix $\oro=\sum_{i=1}^N|\az_i\>\<\az_i|$.
The BV variational principle is usually applied at the mean-field level where the variational space of $\oD$ 
is  restricted to independent particle states, i.e., with $\oD=\kfi\bfi$.

\subsection{Expectation values of one-body operators}

In addition to the mean-field approximation, the variational space for $\oB$ is usually constrained to belong to the same class of operators as the observable of interest. 
For instance, if one wants to predict expectation values of one-body observables $\oX=\sum_{i=1}^N\oq_X(i)$, 
then it is natural to restrict the variational space for $\oB$ to one-body operators. 
In this case, one recovers the TDHF equation~\cite{dir30,bal85}
\oeq 
i\hbar\frac{\partial \ro}{\partial t}=\[h[\ro],\ro\],
\label{eq:tdhf}
\ceq
where $h[\ro]$ is the Hartree-Fock (HF) single-particle Hamiltonian with matrix elements $h_{\al\be}=\frac{\delta \bfi \oH \kfi}{\delta \ro_{\be\al}}$, $\oH$ is the full Hamiltonian, and $\ro_{\al\be}=\<\az_\al|\oro|\az_\be\>=\bfi \oad_\be\oa_\al \kfi$.

According to this variational approach, TDHF is an optimized mean-field theory 
to describe expectation values of one-body observables.
However, TDHF
may fail to reproduce their fluctuations $\sigma_{XX}=\sqrt{\<\oX^2\>-\<\oX\>^2}$~\cite{koo77,das79}.

\subsection{Fluctuations of one-body operators}

The BV variational principle can also be used with the variational space $\oB\in\{e^{\ga\oad\oa}\}$
to determine an optimum mean-field prediction for correlations $\sigma_{XY}$ and fluctuations $\sigma_{XX}$ 
of one-body operators~\cite{bal84,bal92}, 
with
\oeq
\sigma_{XY}=\sqrt{\<\oX\oY\>-\<\oX\>\<\oY\>}.
\label{eq:corr_def}
\ceq
In case of independent particle states, this leads to
\oeq
\sigma_{XY}^2(t_1)=\lim_{\epsilon\rightarrow0}\frac{1}{2\epsilon^2}\tr \(\[\ro(t_0)-\ro_X(t_0,\epsilon)\]\[\ro(t_0)-\ro_Y(t_0,\epsilon)\]\) ,
\label{eq:corr}
\ceq
where $\tr$ denotes a trace in the single-particle space. 
The one-body density matrices $\ro_X(t,\epsilon)$ obey 
the TDHF equation~(\ref{eq:tdhf}) with the boundary condition 
\oeq
\ro_X(t_1,\epsilon)=e^{i\epsilon q_X}\ro(t_1)e^{-i\epsilon q_X},
\ceq
while $\ro(t)$ is the solution of Eq.~(\ref{eq:tdhf}) with the initial condition 
$\ro_{\al\be}(t_0)=\Tr\oad_\be\oa_\al\oD_0=\<\phi_0| \oad_\be\oa_\al |\phi_0\>$.
The optimum mean-field prediction of $\sigma_{XY}$ in Eq.~(\ref{eq:corr}) differs from the ''standard'' TDHF expression 
which is evaluated from Eq.~(\ref{eq:corr_def}) using $\ro(t_1)$.

Eq.~(\ref{eq:corr}) has been solved numerically in the past with simple effective interactions and geometry restrictions~\cite{mar85,bon85,tro85}.
Modern three-dimensional TDHF codes with full Skyrme functionals~\cite{kim97,nak05,mar05,uma05}
can now be used for realistic applications of the BV variational principle~\cite{bro08,sim11}. 

In this work, the fluctuations $\sigma_{NN}$, $\sigma_{ZZ}$, and $\sigma_{AA}$, 
are computed in fragments resulting from actinide collisions.
The correlations $\sigma_{NZ}$, which are strictly zero in standard TDHF calculations, are also determined with this approach.

\section{numerical details}

The use of a three-dimensional TDHF code
with a full Skyrme energy-density-functional (EDF), modeling nuclear interactions between nucleons and including spin-orbit interaction~\cite{sky56,cha98}, 
allows for a realistic prediction of these quantities. 
The TDHF equation~(\ref{eq:tdhf}) is solved iteratively in time, with a time step~$\Delta{t}=1.5\times10^{-24}$~s.
The single-particle wave-functions are evolved on a Cartesian grid of $96\times32\times32/2$ points 
with a plane of symmetry (the collision plane) 
and a mesh-size $\Delta{x}=0.8$~fm. The initial distance between collision partners is 22.4~fm.
The {\textsc{tdhf3d}} code is used with the SLy4$d$ parameterization~\cite{kim97} of the Skyrme EDF, which
 is the only phenomenological ingredient, 
 as it has been adjusted on nuclear structure properties~\cite{cha98}. 
Ref.~\cite{sim10a} gives more details of the TDHF calculations.
The numerical details for the evaluation of Eq.~(\ref{eq:corr}) can be found in~\cite{sim11}.

\section{Collision time in $^{238}$U+$^{238}$U}

TDHF calculations have been performed to investigate the collision time in $^{238}$U+$^{238}$U~\cite{gol09}.
The $^{238}$U nucleus exhibits a prolate deformation with a symmetry axis in its ground state. 
The effect of this deformation on collision is investigated in four configurations ($xx$, $yx$, $yy$ and $yz$) associated to different initial orientations. 
The letters $x$, $y$ and $z$ denote the orientation of the symmetry axis of the nuclei which collide along the $x$ axis [see 
Fig.~\ref{fig:densities}]. We focus on central collisions as they lead to the most dissipative reactions with the longest collision times.

\begin{figure}
\resizebox{1.0\columnwidth}{!}{
\includegraphics{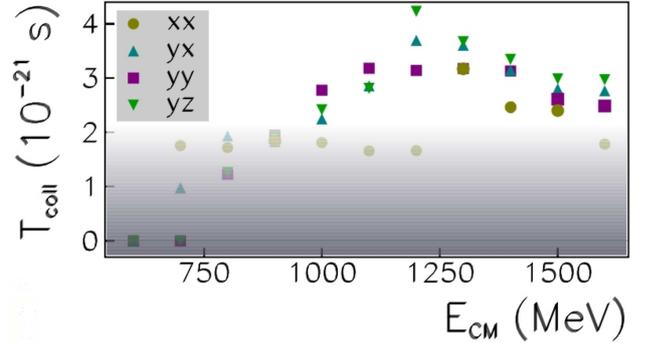}}
\caption{Collision times for each orientation as function of center of mass energy. The shaded area indicates the limit of 2~zs above which vacuum decay is expected to be observable in central collisions.}
\label{fig:transfer}
\end{figure}

Here,   
the collision time $T_{coll}$ is defined as the time during which the neck density exceeds $\rho_0/10$. 
It is shown in Figure~\ref{fig:transfer} as function of the center of mass energy $E_{cm}$.  
At $E_{c.m.}\leq900$~MeV,  three distinct behaviors between the $xx$, $yx$ and $yy/yz$ configurations are seen.  In particular, the last need more energy to get into contact as the 
energy threshold above which nuclear interaction plays a significant role is higher for such compact configurations. 

At all energies, the $yx$, $yy$ and $yz$ orientations exhibit roughly similar behaviors, i.e., a rise and fall of $T_{coll}$ with a maximum of $3-4\times10^{-21}$~s at $E_{c.m.}\sim1200$~MeV. 
Dynamical evolution of nuclear shapes in these three configurations 
and a strong transfer in the~$yx$ one (see next section) 
are responsible for these rather long collision times as compared to scattering with frozen shapes of the reactants~\cite{zag06}. 
The~$xx$ configuration, however, behaves differently.  
For $700<E_{c.m.}<1300$~MeV, $T_{coll}$ exhibits a plateau which does not exceed~$2\times10^{-21}$~s.
This overall reduction of $T_{coll}$ in the~$xx$ case is attributed to the strong overlap of the tips,   
producing a density in the neck higher than $\rho_0$~\cite{gol09}. 
The fact that nuclear matter is difficult to compress translates into a strong repulsive force between the fragments which decreases their contact time. 
This phenomenon is also responsible for the fall of collision times in the other configurations, though 
higher energies 
are needed to strongly overlap.

The calculations of  Ref.~\cite{ack08} show that the observation of spontaneous emission of $e^++e^-$ needs a contact time of at least 2~zs between the bare uranium nuclei. The TDHF calculations of Fig.~\ref{fig:transfer} predict that such contact times are reached in central collisions for energies $E_{cm}>1$~GeV. 
This lower energy limit should be taken into account in future experimental programs dedicated to the search of QED vacuum decay. 
Finally, it is worth mentioning that other approaches lead to comparable collision times in actinide collisions~\cite{tia08,sar09}. 

\section{Formation of neutron-rich transfermium nuclei}

\subsection{ $^{238}$U+$^{238}$U reaction}

We now analyze the proton and neutron numbers of the fragments produced in exit channels of actinide collisions. 
Strictly speaking, these fragments should be considered as {\it primary} fragments as they might decay by statistical fission. 
This decay is not studied here as it occurs on a much longer time scale than the collision itself.
The importance of initial orientation on reaction mechanism is clearly seen in Fig.~\ref{fig:densities} for the $^{238}$U+$^{238}$U reaction. 
For symmetry reasons, the $xx$, $yy$, and $yz$~configurations give two symmetric distributions of fragments, 
 although nucleon transfer is still possible 
thanks to particle number fluctuations.
Nucleon transfer is expected to be stronger in the $yx$ configuration because, in addition to fluctuations, no spatial symmetry prevents from an average flux of nucleons. 
The $yx$ configuration is then expected to favor the formation of nuclei heavier than~$^{238}$U. 

\subsection{ $^{232}$Th+$^{250}$Cf reaction}

Similar calculations have been performed on the system $^{232}$Th+$^{250}$Cf~\cite{ked10}.
The same effect is observed, i.e., an important multi-nucleon transfer in the $xy$ and $yx$ configurations. 
The $xy$ configuration where the $^{250}$Cf nucleus receives nucleons (its deformation axis is perpendicular to the collision axis while the one of $^{232}$Th is parallel to the collision axis) corresponds to an ''inverse quasifission'' mechanism due to a specific orientation of the collision partners.
Indeed, contrary to standard quasifission, the exit channel is more mass asymmetric than the entrance channel. 
Note that inverse quasifission may also occur due to  shell effects in the exit channel~\cite{zag06}. 

\begin{figure*}
\resizebox{2.0\columnwidth}{!}{
\includegraphics{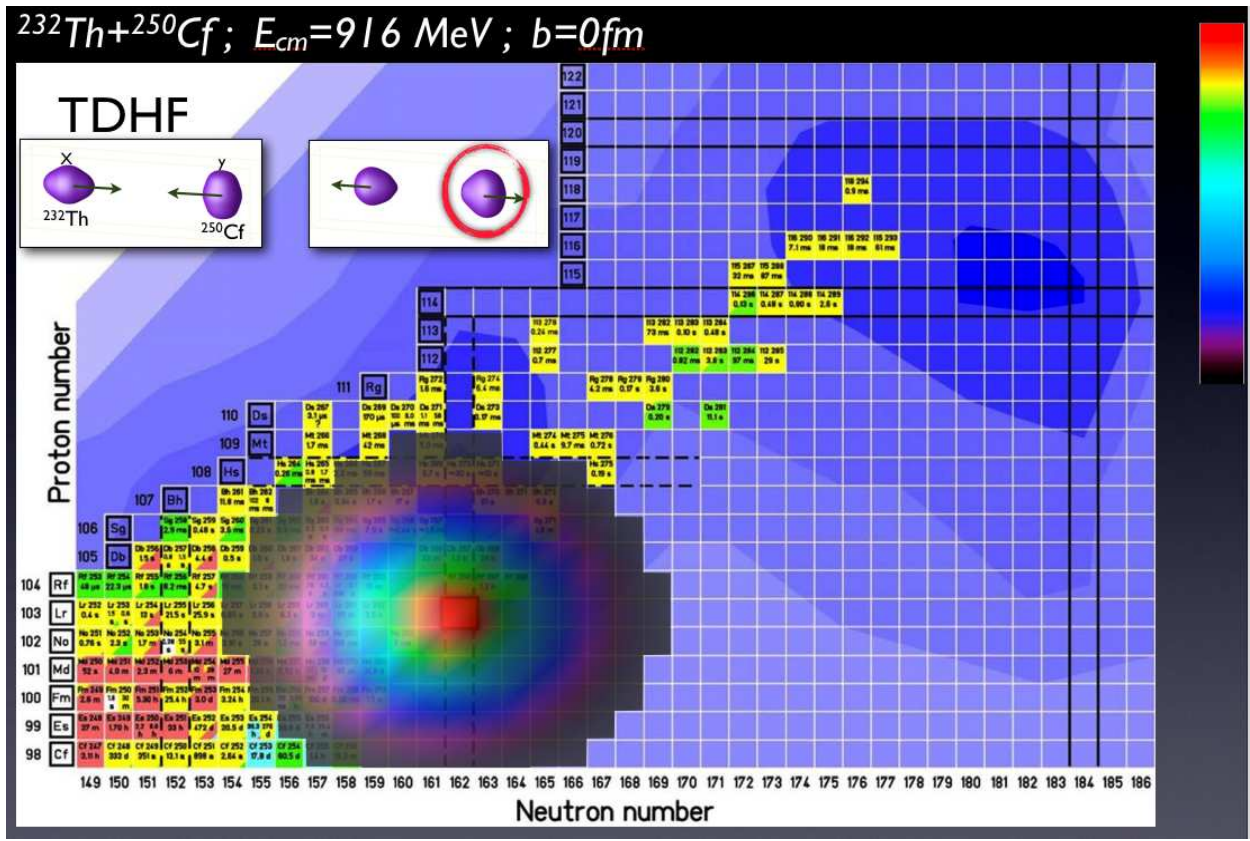}}
\caption{TDHF heavy fragment probability distribution (linear scale) for $^{232}$Th+$^{250}$Cf central collision in the $xy$ configuration.}
\label{fig:Lr_TDHF}
\end{figure*}

The effect is illustrated in Fig.~\ref{fig:Lr_TDHF} where the distribution of heavy fragments is shown at $E_{cm}=916$~MeV.
This distribution is computed after a  TDHF calculation, using a particle number projection technique~\cite{sim10b}.
The center of the distribution is located around $^{265}$Lr, i.e., in the neutron-rich side of the known Lawrencium isotopes. Note that, at the end of the TDHF calculation, the decay of the fragments by neutron emission is only partial~\cite{ked10}.

The width of such a distribution is known to be underestimated at the TDHF level~\cite{koo77,das79}.
In addition, we see in Fig.~\ref{fig:Lr_TDHF} that the probability distributions for $N$ and $Z$ are uncorrelated at the TDHF level. 
This is not a feature of the TDHF formalism itself, but rather a limitation due to the fact that, for practical applications, one assumes $|\phi\>=|\phi_p\>\times |\phi_n\>$, where $|\phi_p\>$ and $|\phi_n\>$ are Slater determinants of the proton and neutron single-particle wave-functions, respectively~\cite{sim11}. 
 If this constraint is released, as in~\cite{sim05}, then non-zero correlations could be obtained at the TDHF level. 

\begin{figure*}
\resizebox{2.0\columnwidth}{!}{
\includegraphics{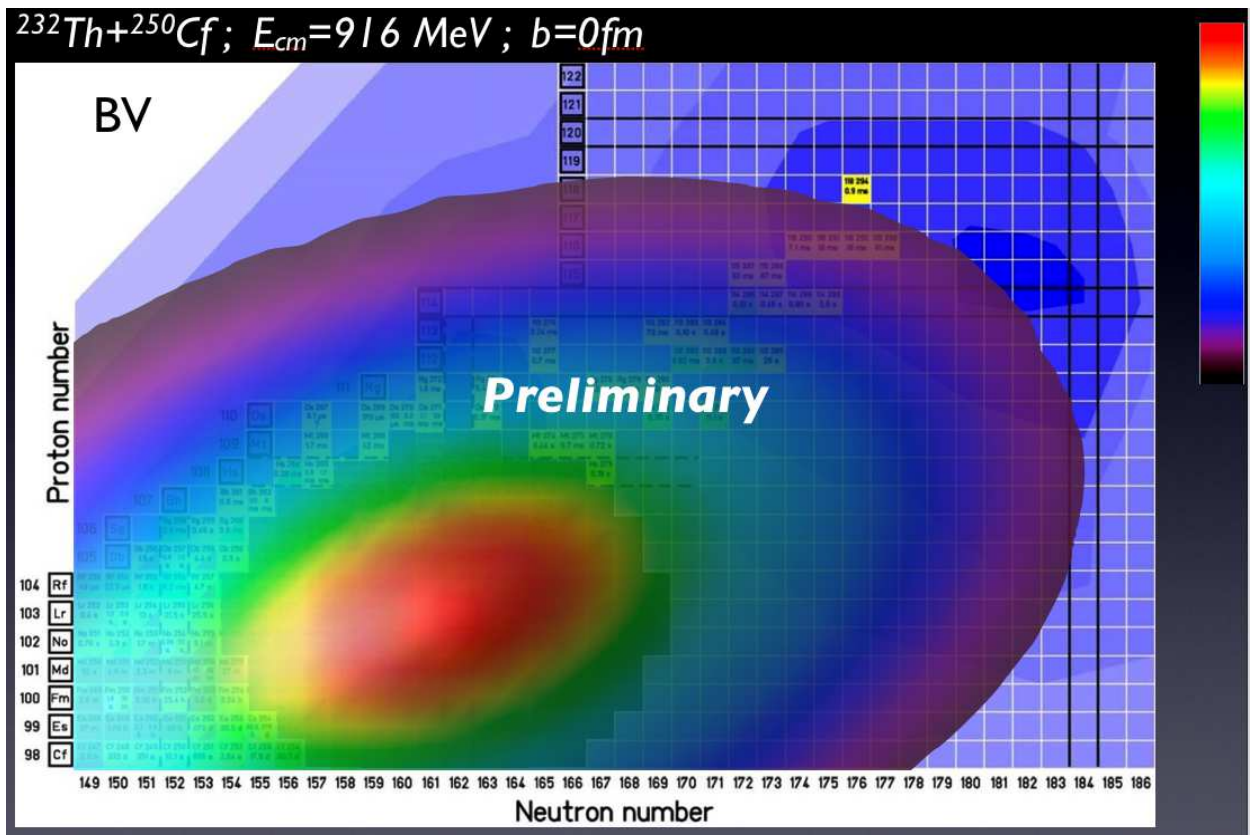}}
\caption{Same as Fig.~\ref{fig:Lr_TDHF} with the BV prescription.}
\label{fig:Lr_BV}
\end{figure*}

The Balian-V\'en\'eroni variational principle can be used, at the mean-field level, to optimize both widths of proton and neutron distributions as well as their correlations. Realistic calculations have been performed recently to study deep-inelastic collisions~\cite{sim11}. 
Similar calculations have been done to investigate the inverse quasifission mechanism discussed above. 
Preliminary results are shown in Fig.~\ref{fig:Lr_BV} where the heavy-fragment distribution is shown for the same $xy$ orientation of $^{232}$Th+$^{250}$Cf as in Fig.~\ref{fig:Lr_TDHF}. As expected, much larger widths than in the TDHF case are observed in one hand, and, in the other hand, strong correlations between the proton and neutron number distributions are observed, which can be seen by the fact that the fragments are produced along the valley of stability.

\section{Conclusions}

To conclude, this fully microscopic quantum investigation of 
actinide collisions exhibits a rich phenomenology 
strongly influenced by the shape  of the 
nuclei.  
Two main conclusions can be drawn. 
($i$) The giant system formed in bare uranium-uranium  central collisions is expected to 
survive enough time with an energy $E_{c.m.}\geq1000$~MeV, for the spontaneous positron emission to occur. 
($ii$) The primary heavy-fragments produced by multinucleon transfer are more neutron-rich than in fusion-evaporation reactions. 
The width of these distributions, computed with the Balian-V\'en\'eroni prescription, are much larger than with TDHF. Associated cross-sections need to be determined to estimate the experimental possibility of neutron-rich transfermium and SHE productions. 
Extension of the formalism need to be investigated.
For instance, the role of pairing could be studied with Time-Dependent Hartree-Fock-Bogoliubov (or BCS) codes~\cite{ave08,eba10,was10}. 
Stochastic-mean-field methods might also be applied to 
investigate the role of initial beyond-mean-field correlations on fluctuations~\cite{ayi08}.

\section{Acknowledgements}

We thank P. Bonche for providing his code.
We are also grateful to M. Dasgupta, D. J. Hinde, and D. Lacroix for useful discussions. 
CS and DJK thank the Australian National University in which part of this work has been done.
The calculations have been performed in the 
Centre de Calcul Recherche et Technologie of the Commissariat \`a l'\'Energie Atomique
and on the NCI National Facility
in Canberra, Australia, which is supported by the Australian Commonwealth Government.
Support from ARC Discovery grants DP0879679 and DP110102879 is acknowledged.

%

%
%


\end{document}